\documentclass[a4paper,fleqn,usenatbib]{mnras}

\usepackage{newtxtext,newtxmath}

\usepackage[T1]{fontenc}
\usepackage{ae,aecompl}

\usepackage{graphicx}	
\usepackage{amsmath}	
\usepackage{amssymb}	

\defcitealias{SLC2015}{SLC15}
\newcommand{\galaxy}{J0403}

\title[A lightweight galaxy lens at z=0.066]{A new strong-lensing galaxy at z=0.066: Another elliptical galaxy with a lightweight IMF}

\author[Collier W. et al.]{
William P. Collier,\thanks{E-mail: william.p.collier@durham.ac.uk}
Russell J. Smith,
John R. Lucey
\\
Centre for Extragalactic Astronomy, Departments of Physics, University of Durham, Durham DH1 3LE, UK\\
}

\date{Submiteed to MNRAS 16th March 2018}

\pubyear{2018}

\begin{document}
\label{firstpage}
\pagerange{\pageref{firstpage}--\pageref{lastpage}}
\maketitle

\begin{abstract}

We report the discovery of a new low-redshift galaxy-scale gravitational lens, identified from a systematic search of publicly available MUSE observations. The lens galaxy, 2MASXJ04035024-0239275, is a giant elliptical at $z$\,=\,0.06604 with a velocity dispersion of $\sigma$\,=\,314\,km\,s$^{-1}$.  The lensed source has a redshift of 0.19165 and forms a pair of bright images either side of the lens centre.  The Einstein radius is 1.5\,arcsec, projecting to 1.8\,kpc, which is just one quarter of the galaxy effective radius. After correcting for an estimated 19 per cent dark matter contribution, we find that the stellar mass-to-light ratio from lensing is consistent with that expected for a Milky Way initial mass function (IMF). Combining the new system with three previously-studied low-redshift lenses of similar $\sigma$, the derived mean mass excess factor (relative to a Kroupa IMF) is $\langle\alpha\rangle$\,=\,1.09$\pm$0.08. With all four systems, the intrinsic scatter in $\alpha$ for massive elliptical galaxies can be limited to $<0.32$, at 90 per cent confidence.

\end{abstract}





\section{Introduction}
\label{sec:Intro}

The IMF is fundamental to understanding galaxy formation and evolution, as well as to interpreting observed properties (e.g. estimating stellar masses). Within the different star forming environments in the Milky Way, the stellar initial mass function (IMF) is well constrained, and approximately invariant \citep[][]{Bastian2010,Offner2014}. Deviations toward a flatter IMF have been reported for some resolved ultra-faint Local Group dwarfs \citep{Geha2013}.


For galaxies beyond the Local Group, however, resolved studies are not possible, and the IMF must be inferred from the integrated light and/or gravitational mass tracers. Broadly the observational techniques fall into two categories. The first method infers the stellar population via high signal-to-noise spectroscopy. The strength of gravity sensitive absorption lines are measured by fitting detailed stellar population synthesis templates. The second method indirectly measures the stellar population by comparing a stellar mass-to-light ratio ($M/L$) measured from stellar dynamics or strong lensing, to a reference $M/L$ from a fixed IMF stellar population model. Studies utilising both techniques independently have found for early-type galaxies (ETGs), have found evidence for an increasingly `heavy' IMF (more measured mass than a fixed IMF model predicts) in the most massive ETGs \citep[e.g.][]{Treu2010b,Cappellari2012,Conroy2012b,LaBarbera2013}. 

The mass measurements from gravitational lensing require careful treatment to disentangle the contributions of stellar mass and dark matter (DM). For example, \citet{Treu2010b} analysed lenses at $z$\,$\sim$\,0.2 by combining lensing and stellar kinematics to constrain the parameters of a two-component mass model. This approach involves several assumptions (spherical geometry, constant stellar mass-to-light ratio, etc). In contrast, {\it nearby} ($z$\,$\leq$\,0.1) lenses offer a geometry in which the Einstein radius ($R_{\rm Ein}$) is reached at smaller physical radii and hence probes the dense stellar-dominated core. In such cases the relative uncertainty from DM is minimized, as the ratio of dark to stellar matter is reduced and ``pure'' lensing constraints on the stellar mass can be obtained.



At present, the best studied $z$\,$<$\,0.1 lensing ellipticals are those discovered from the SINFONI Nearby Elliptical Lens Locator Survey (SNELLS) \citep*{SLC2015}. This targeted approach used an integral field unit (IFU) to search for background emitters behind massive elliptical galaxies, finding SNL-1 ($z$\,=\,$0.031$) and SNL-2 ($z$\,=\,$0.052$), and ``rediscovering'' the previously-known lens SNL-0 ($z$\,=\,$0.034$) \citep{Smith2005,Smith2013}. The SNELLS galaxies yielded lensing masses in strong disagreement with `heavy' IMFs for massive ETGs, instead measuring $M/L$ consistent with a Milky-Way (MW) like IMF \citep{Kroupa2001}, both in low resolution ground based \citep{SLC2015}, and in high resolution space-based observations \citep*{Collier2018}. 

With only three galaxies, the possibility that the SNELLS sample \textit{by chance} was drawn from the ``tail'' of an intrinsically broad distribution in IMFs cannot be ruled out. As such, with low number statistics, increasing the sample size is essential to investigate further this conclusion,  and to test whether these lenses are representative of the parent population.

Discovering low-redshift lenses is challenging due to the high surface brightness of the massive foreground galaxy. While SNELLS pre-selected high-probability lenses using velocity dispersion measurements, an alternative approach is to exploit large multi-IFU surveys like SAMI and MaNGA \citep{Bryant2015,Bundy2015}, which have recently yielded several new systems \citep{Smith2017,Talbot2018}. A third technique, which we are currently pursuing, is to search for lensed line emitters behind galaxies targeted for other science goals, using data from public archives. Observations made with MUSE (Multi-Unit Spectroscopic Explorer) \citep{Bacon2014} on the ESO Very Large Telescope are well suited to this method, due to the very high sensitivity and wide spectral range of the instrument.

In this paper, we present the discovery of the first multiply-imaged strong gravitational lens from this programme. In Section \ref{sec:data}, we briefly describe the modified lens search process. In Section \ref{sec:lensdiscovery} we present the new lens system properties. In Section \ref{sec:plensing} we present lensing mass constraints and the IMF mass excess parameter, and then compare, and combine our results with the SNELLS sample in Section \ref{sec:conclusion}.



\section{A MUSE archival lens search}
\label{sec:data}

Our new lensing system was identified in the course of a systematic search for multiply-imaged line emitters behind low-redshift ETGs in public archival MUSE data.  A full description of this programme will be presented elsewhere (Collier et al., in prep); here, we briefly summarise the sample selection and main processing steps, to provide context for the new discovery.

For this lens search, we are analysing public MUSE observations which overlap with luminous nearby galaxies selected from the 2MASS Redshift Survey \citep{Huchra2012}. We focus our search on targets with redshift 5\,000\,$<$\,$cz$\,$<$\,20\,000\,km\,s$^{-1}$, and $M_K$\,$<$\,--25.7. 

The analysis begins with ``Phase 3'' pipeline-reduced products retrieved from the ESO Science Archive. Our primary goal is to detect faint lensed background emitters behind the central regions of nearby ETGs. Before attempting line detection, we therefore process the pipeline-reduced data-cubes to subtract the bright stellar foreground, effectively fitting elliptical isophote models to each wavelength channel. The residual cubes are noise-normalised and filtered to suppress instrumental artifacts. 

Candidate emission lines are then identified above a significance threshold in the processed residual cubes. Spectra for likely sources are extracted, and redshifts are estimated automatically using {\sc marz} \citep{Hinton2016}. Finally, the resulting object catalogue is visually inspected, and searched for any spatially-separated line-emitters with similar redshifts, which are potentially multiply-imaged background galaxies.

Although we optimise our search method to detect faint background emission, the first galaxy-scale lens discovered was over 100 times brighter than our detection limit. Hence we report this system as a ``special case'' in this paper.

\section{The new lens}

\label{sec:lensdiscovery}

2MASXJ04035024--0239275 (hereafter \galaxy{}) is a luminous ($M_K$\,=\,--25.7) elliptical galaxy at heliocentric redshift $z_{\rm H} = 0.06655$ \citep{Jones2009}. In imaging from the Pan-STARRS PS1 survey \citep{Chambers2016} J0403 shows a smooth inner light profile, well fitted by a Sersic profile with index $n$\,$\approx$\,4. In the outer regions (radius $>$\,20\,arcsec), some low-surface-brightness ($i$--band $\sim$25 mag arcsec$^{-2}$) tidal structures are visible. \galaxy{} is a relatively isolated galaxy, with no overdensity of similar-colour objects visible in PS1 imaging. The closest comparable brightness companion (2MASXJ04034531--0236595) is a spiral galaxy, fainter by 1.3 magnitudes, with relative velocity +550\,km\,s$^{-1}$, at a projected distance of $\sim$\,210\,kpc (2.7\,arcmin). There are no catalogued clusters with comparable redshift within a radius of 2\,degrees (9.5\,Mpc). 

\galaxy\ was observed with MUSE as part of a study targeting supernova hosts (ESO programme 098.D-0115(A); PI L. Galbany). The exposure time was $\sim$\,2400s, with seeing $\lesssim$\,0.8 arcsec (estimated FWHM from compact sources). The MUSE field-of-view is 1$\times$1\,arcmin$^2$ ($\sim$\,80$\times$80\,kpc$^2$).

\begin{figure*}
	\includegraphics[width=1.95\columnwidth]{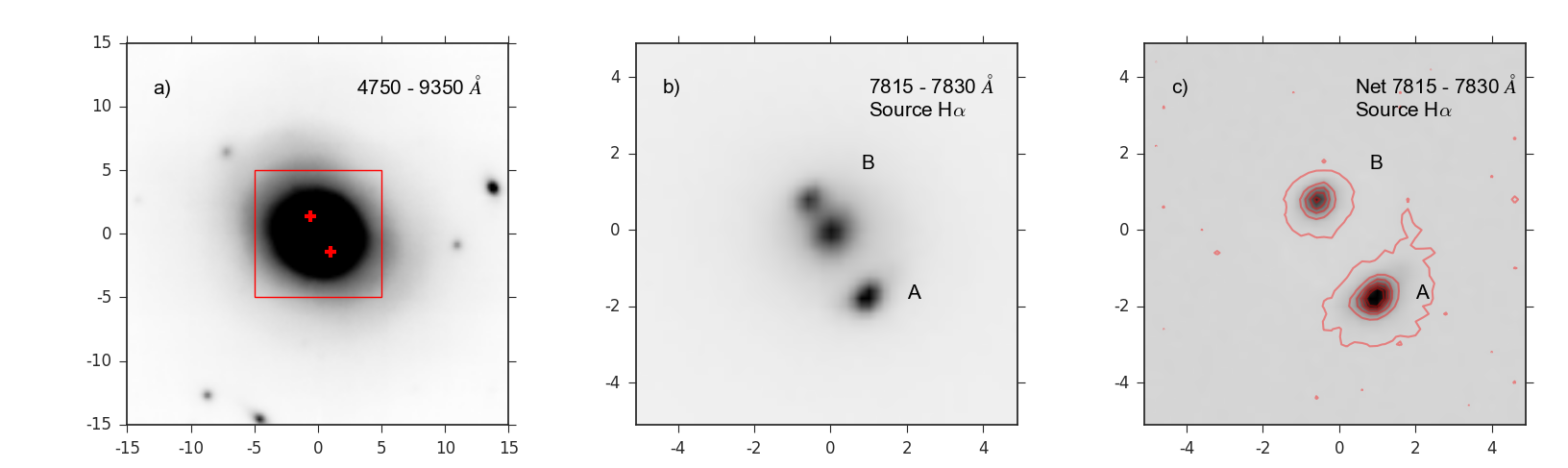}
	\caption{MUSE data of \galaxy{}, within different wavelength ranges (shown in the top right). a) displays a broadband image collapsed over the entire MUSE wavelength range, displaying the lens structure. The red crosses mark the arc positions, while the red box encompasses the region shown in panels b) and c). In b) we collapse about the H\,$\alpha$ line in the background source, revealing arcs A and B, separated by 2.94\,$\pm$\,0.06 arcsec, without the need for lens subtraction. c) shows the continuum-subtracted image at the same wavelength, with contours to show the outer isophotes of the arcs.}
	\label{fig:MUSEspatial}
\end{figure*}

\begin{figure*}
	\includegraphics[width=2.0\columnwidth]{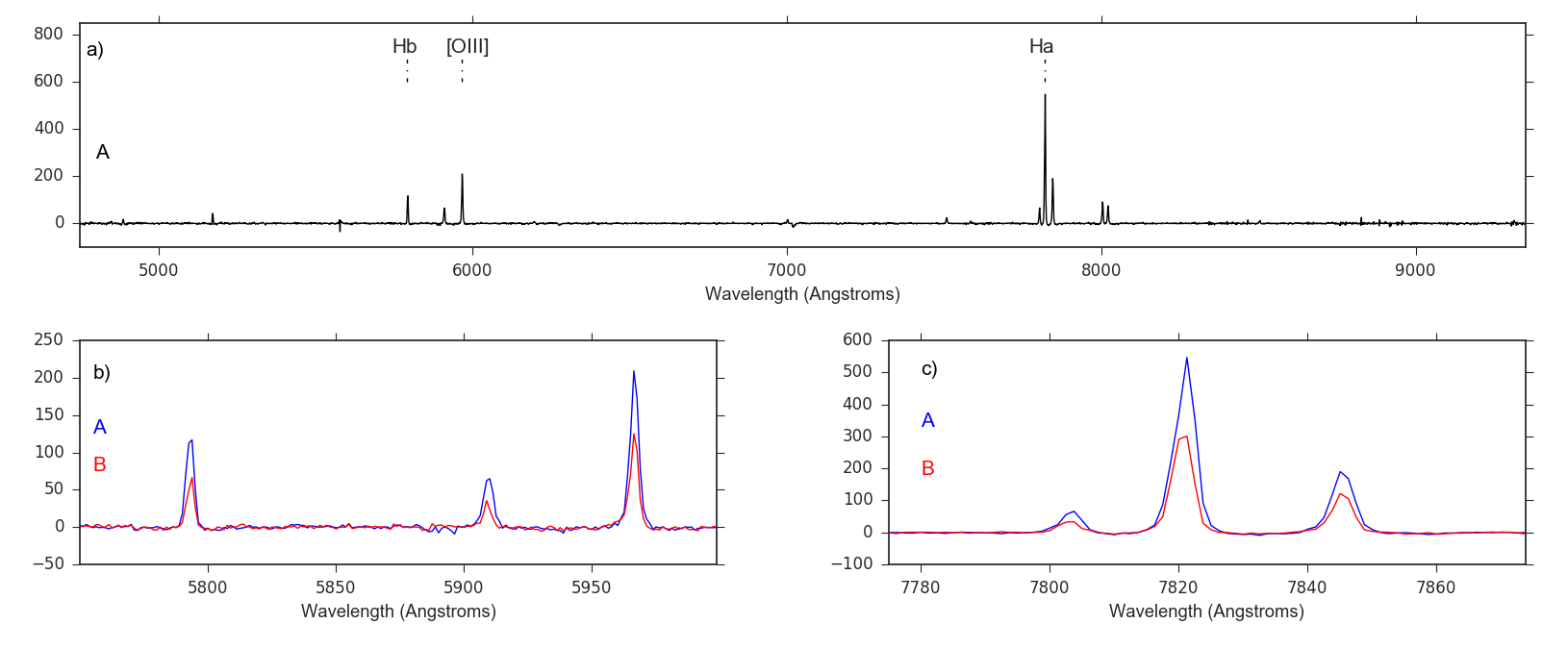}
	\caption{MUSE spectra of the lensed images. In a) we display the arc spectrum extracted from a continuum-subtracted residual datacube, showing the bright H$\alpha$, [N\,\textsc{ii}], [O\,\textsc{iii}], H\,$\beta$, H\,$\gamma$ and [S\,\textsc{ii}] emission lines, at a redshift of 0.19165. In b) and c), we overlay the emission from images A and B, for the \textsc{[Oiii]} and H$\alpha$ regions, respectively. There is negligible velocity offset ($\la$\,50 km\,s$^{-1}$) between the two spectra, and image A has been subject to a greater magnification. }
	\label{fig:MUSEspectral}
\end{figure*}

In the MUSE data (Figure \ref{fig:MUSEspatial}a), the light profile of \galaxy\ is smooth at small radii, with evidence for the faint tidal features seen in the PS1 imaging. Extracting narrow band images (Figure \ref{fig:MUSEspatial}b) highlights a pair of extended background emitters (arcs A and B) either side of the galaxy centre. Strong H\,$\alpha$, [O\,\textsc{iii}], and H\,$\beta$ lines are present in the spectra of both emitters, with the same redshift, $z_{\rm H}$\,=\,0.19165 (Figure \ref{fig:MUSEspectral}). The respective velocity offset between A and B is $\leq$\,50\,km\,s$^{-1}$ providing strong evidence of a common source, and confirming this as a lens system. When the continuum is subtracted, image A shows slight curvature at low surface brightness levels (Figure \ref{fig:MUSEspatial}c). There is some evidence of a small velocity gradient ($\sim$\,50 km\,s$^{-1}$ peak-to-peak) across arc A, suggestive of a face-on or low mass source. 

\begin{figure*}
	\centering
	\includegraphics[width=0.98\linewidth]{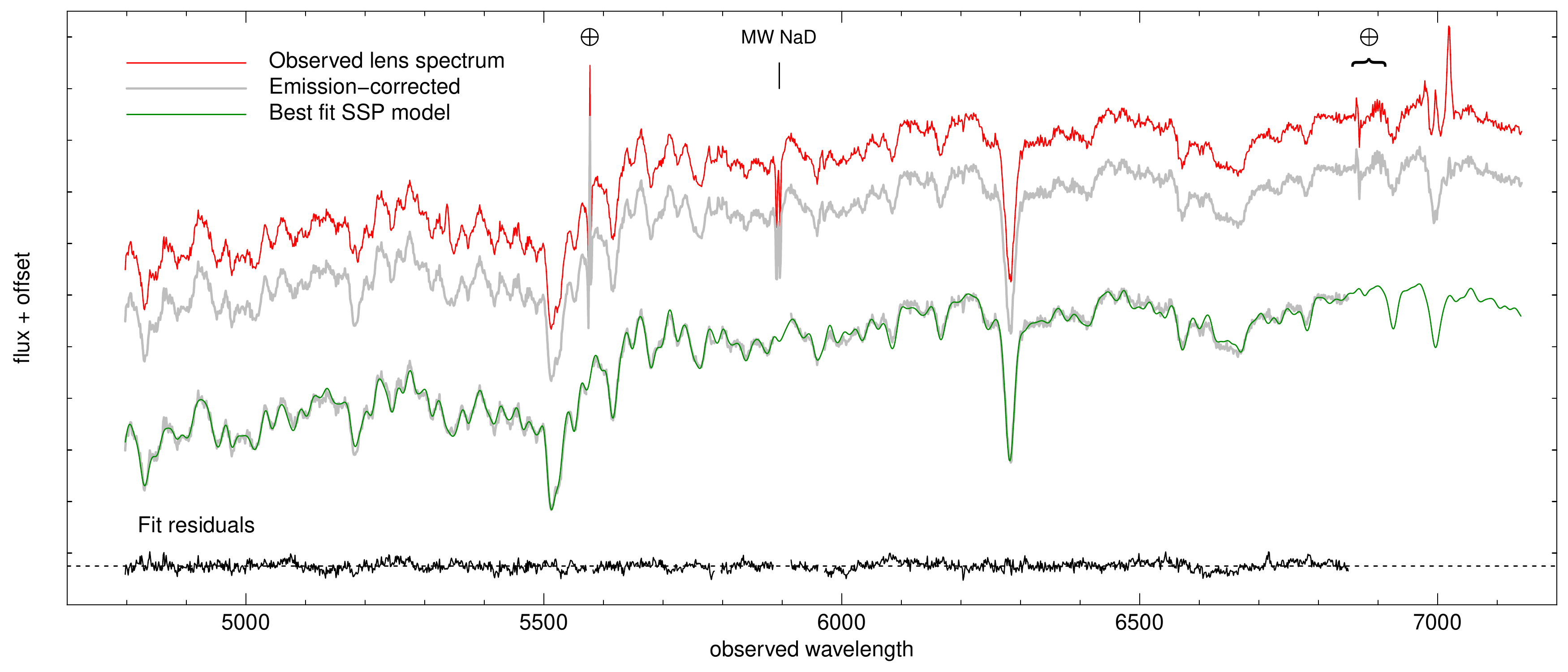}
	\caption{The MUSE spectrum of the lens galaxy \galaxy, extracted within the Einstein aperture, i.e. radius 1.5\,arcsec, after masking pixels strongly affected by the arcs.  For clarity, only the blue region used for fitting stellar population models is shown. The observed spectrum (red) shows that the galaxy has an absorption-dominated spectrum, but also has a nebular line emission component, seen most easily in the H$\alpha$--[N\,{\sc ii}] region. The corrected spectrum, shown in grey, after subtracting an emission-line model, fitted to the H$\alpha$--[N\,{\sc ii}] complex, and assuming Case B recombination and galactic (no internal) extinction, to predict the H$\beta$ emission. Below, we reproduce the emission-corrected spectrum, and show the best-fitting stellar population (green), derived from the models of \citet{Conroy2012}. This model has age 12\,Gyr and metal abundances typical for massive ellipticals. The fit residuals, shown in black, have a 1 per cent rms. In fitting this model, we exclude the H$\alpha$ region, as well as wavelengths contaminated by atmospheric artifacts (indicated by \earth) and the MW neutral sodium doublet.}
	\label{fig:lsqlensspec}
\end{figure*}

The lensed images are separated by 2.94\,$\pm$\,0.06\,arcsec ($\sim$\,3.75\,kpc), where the uncertainty is derived from measurements at various emission lines. For our lensing analysis we adopt $R_{\rm Ein}$ as half of the image separation, 1.47\,$\pm$\,0.03\,arcsec, and extract an aperture spectrum ($\leq$\,$R_{\rm Ein}$) of \galaxy{}, shown in Figure \ref{fig:lsqlensspec}. The spectrum displays the strong absorption features typical of an old stellar population, with strong Na\,D, Mg I and TiO, along with weak H\,$\alpha$ and H\,$\beta$ emission. The measured heliocentric redshift, $z_{\rm H}$\,=\,0.06604, and velocity dispersion of $\sigma$ = 314\,$\pm$\,5 km s$^{-1}$, are found by fitting this central region using \textsc{ppxf} \citep{Cappellari2004}.

Using $i$-band imaging from the Pan--STARRS (PS1) survey we measure a total magnitude for J0403 of 14.26\,$\pm$\,0.04 mag and an effective (half-light) radius of 5.7\,arcsec (i.e. 3.9 times the size of $R_{\rm Ein}$). Within $R_{\rm Ein}$ we measure an aperture magnitude of 16.35\,$\pm$\,0.01 mag. As the PS1 point spread function (PSF)($\sim$\,1.1 arcsec) is comparable in size to $R_{\rm Ein}$, a PSF correction is necessary. We determine the correction by fitting a Sersic model via {\sc galfit} \citep{Peng2010}, with a convolution kernel derived from a set of adjacent stars. The magnitude difference between the Sersic model before and after convolution was 0.22\,$\pm$\,0.02 mag and hence we adopt a PSF-corrected Einstein-aperture magnitude of $i_{\rm Ein}$\,=\,16.13\,$\pm$\,0.02 mag.

\section{Lensing mass and the IMF}
\label{sec:plensing}

Gravitational lensing provides a precise measurement of the total projected mass within the Einstein radius ($M_{\rm Ein}$). Measuring $M_{\rm Ein}$ and the luminosity within the same aperture, ($L_{\rm Ein}$), the resulting mass to light ratio can be related to the IMF mass excess parameter, $\alpha$. This factor is defined as:
\[ 
$$ \alpha = \dfrac{\Upsilon}{\Upsilon_{\rm Ref}}  =  \dfrac{M^*_{\rm Ein}}{ L_{\rm Ein}} \times \dfrac{1}{\Upsilon_{\rm Ref}} = \dfrac{M_{\rm Ein} - M^{\rm DM}_{\rm Ein}}{\Upsilon_{\rm Ref} L_{\rm Ein}} $$  \ \ , 
\]
where $M^{\rm DM}_{\rm Ein}$ is the dark matter component. We compare $\Upsilon$, the observed stellar mass-to-light ratio, to $\Upsilon_{\rm Ref}$, a reference mass-to-light ratio for a modelled stellar population with comparable properties (i.e. metallicity, age), with a fixed \citet{Kroupa2001} IMF. Hence a Kroupa IMF has $\alpha$\,=\,1 by definition, while a Salpeter IMF \citep{Salpeter1955}, with more low mass stars, has $\alpha$\,=\,1.55.

We compute $M_{\rm Ein}$ using the symmetric lens equation \citep[see][section 4.1]{SLC2015}. We adopt cosmological parameters from the 7-year \textit{Wilkinson Microwave Anisotropy Probe} (WMAP), i.e. H$_0$ = 70.4 km s$^{-1}$ Mpc$^{-1}$, $\Omega _{\rm m}$ = 0.272 and $\Omega_{\Lambda}$ = 0.728 \citep{Komatsu2011}, to calculate the lensing geometry for redshifts in the CMB frame ($z_{\rm CMB}$\,=\,0.06569,\,0.19130 and $\dfrac{D_{\rm l} D_{\rm s}}{D_{\rm ls}}$\,=\,400.5\,Mpc). We derive a total projected mass, $M_{\rm Ein}$ = 10.64$\pm$0.23\,$\times$\,10$^{10}$ M$_{\odot}$, with the 2 per cent uncertainty dominated by the measurement of $R_{\rm Ein}$. Including a small ellipticity ($e$\,$\approx$\,0.1 as measured from \textsc{galfit}) increases the mass by 3 per cent, and the inclusion of a small external shear ($<$ 5/1 per cent for SIS/SIE) reproduces the image positions perfectly. In order to account for these additional possible complexities, we revise the uncertainty in $M_{\rm Ein}$ to $\sim$4 per cent, i.e. 0.4\,$\times$\,10$^{10}$\,M$_{\odot}$.

The DM mass component is estimated following \citet{SLC2015}, using the \textsc{eagle} hydrodynamical cosmological simulation \citep{Schaye2015}. We measure the average DM mass which would be projected inside an aperture of 1.8\,kpc, averaged over all {\sc eagle} halos hosting galaxies with stellar velocity dispersions $>$275\,km\,s$^{-1}$. This indicates a contribution of M$_{\rm Ein}^{\rm DM}$\,=\,2.01\,$\pm$\,0.36\,$\times$\,10$^{10}$\,M$_{\odot}$ (i.e. 19\% of M$_{\rm Ein}$), which yields an aperture stellar mass $M^*_{\rm Ein}$ of 8.63\,$\pm$\,0.54\,$\times$\,10$^{10}$\,M$_{\odot}$.


The lens aperture luminosity is calculated from the PSF corrected $i_{\rm Ein}$, with additional corrections for the line-of-sight galactic extinction and the \textit{k}--correction. 

 Despite being located at fairly high galactic latitude ($b$\,=\,--38$^\circ$), \galaxy\ lies in a region of relatively high galactic extinction, with $A_i$\,=\,0.26 according to the \citet{Schlafly2011}. Alternatively, maps based on PS1 stellar photometry \citep{Schlafly2014,Green2018} indicate slightly smaller values, with $A_i$\,=\,0.22. The MUSE spectrum shows a clear galactic Na\,D absorption doublet ($\sim$1\,\AA\ equivalent width), which supports the presence of substantial interstellar material along this sightline. We adopt a correction of 0.24\,mag, and allocate an error of 0.04\,mag (16 per cent of the extinction in magnitudes, following \citet{Schlegel1998b}. The \textit{k}--correction is estimated from the lens \textit{g}--\textit{i} colour index (1.7) to be \textit{k}$_i$\,=\,0.04\,$\pm$\,0.01 mag \citep{Chilingarian2010,Chilingarian2012} .

The final corrected Einstein-aperture apparent magnitude is $i_{\rm Ein}$\,=\,15.87\,$\pm$\,0.05, which for the adopted cosmology ($D_{\rm L}$\,=\,295.7\,Mpc), and the solar AB $i$--band absolute magnitude of 4.534 \citep{Blanton2007}, yields an aperture luminosity of 2.59\,$\pm$\,0.12 $\times$10$^{10}$\,L$_{\odot}$. The uncertainty is dominated by the applied corrections. Combined with $M^*_{\rm Ein}$ the observed stellar mass-to-light ratio is $\Upsilon$\,=\,3.33\,$\pm$\,0.26 solar units. 

To convert from mass-to-light ratio ($\Upsilon$) to the IMF mass factor $\alpha$, we need an estimate for the reference mass-to-light ratio for a given fiducial IMF \citep[here][]{Kroupa2001}, which depends on the stellar population properties: age (or star-formation history), metallicity, etc. To estimate these parameters, we make use of the MUSE spectrum of the lens galaxy extracted within the Einstein aperture, shown in Figure \ref{fig:lsqlensspec} (red line). However, the presence of significant nebular emission, combined with limited spectral coverage in the blue, presents some challenges. Specifically, the stellar population age is constrained only by the H$\alpha$ and H$\beta$ absorption lines, but both are contaminated by gas emission, to different extent. 

We tackle the nebular emission by first fitting a three-gaussian model to the H$\alpha$ and [N\,{\sc ii}] lines, with an assumed
stellar continuum model. The relative amplitudes of the  [N\,{\sc ii}] lines are fixed, and all lines are assumed to 
have the same velocity and width. The amplitudes of H$\alpha$ and the [N\,{\sc ii}] doublet are allowed to vary. From this model, we can predict the H$\beta$ emission line, assuming Case B recombination for the intrinsic line ratio, 
and a relative attenuation factor of 1.135 for H$\beta$ from the MW extinction (we assume no internal extinction for this
exercise). Because the underlying H$\alpha$ absorption has only low sensitivity to the stellar population age, 
the emission correction is quite robust against changes to the age assumed in the first step. 

After making this correction to remove the emission-infilling of the spectrum at H$\beta$, we perform a full-spectrum
fit over the interval 4500--6400\,\AA, using single-burst models from \citet{Conroy2012b}, (Figure \ref{fig:lsqlensspec}, green). We allow for variation in abundances of Mg, Fe, Na, and C, as well as variation in age. Formally, the fit implies an old population, with age $12.0\pm1.0$\,Gyr, high metallicity [Z/H]\,$\approx$\,[Mg/H]\,$\approx$\,+0.15$\pm$0.02 and typical massive elliptical galaxy abundance ratios [Mg/Fe]\,$\approx$\,+0.3, [Na/Fe]\,$\approx$\,+0.5, [C/Fe]\,$\approx$\,+0.2.  

The reference mass-to-light ratio is estimated with \citet{Conroy2009} stellar population models accessed via \textsc{ezgal} \citep{Mancone2012}. For a \citet{Kroupa2001} IMF, with an old, solar metallicity stellar population ($z_{\rm form}$\,=\,3), the models predict $\Upsilon^{\rm *}_{\rm Ref}$\,=\,2.90\,$\pm$\,0.10. The uncertainty is derived from small variations in the metallicity and age (e.g. $z_{\rm form}$\,=\,2.5--3.5, metallicity 1--1.5 solar). The resulting IMF mass excess parameter, with its statistical error is $\alpha$\,=\,1.15\,$\pm$\,0.10. 

Several adopted parameters and assumptions affect the value of $\alpha$ derived above.
If we had prescribed the \citet{Planck2016} cosmology, $\alpha$ would be 4 per cent lower, due to the variation in H$_0$. If we had attributed the total lensing mass entirely to the stellar component (i.e. no DM inside $R_{\rm Ein}$), we would have found $\alpha$\,=\,1.42, a 25 per cent increase. If we had assumed, due to the limited blue (most age sensitive) coverage from MUSE, that the population may be 8.5\,Gyrs old ($z_{\rm form}$\,=\,1.5), $\alpha$ would have increased by 14 per cent. If we had adopted the \citet{Maraston2005} models, $\alpha$ would increase by 10 per cent. 

\section{Discussion and Conclusions}
\label{sec:conclusion}

\begin{figure}
	\centering
	\includegraphics[width=1.0\linewidth]{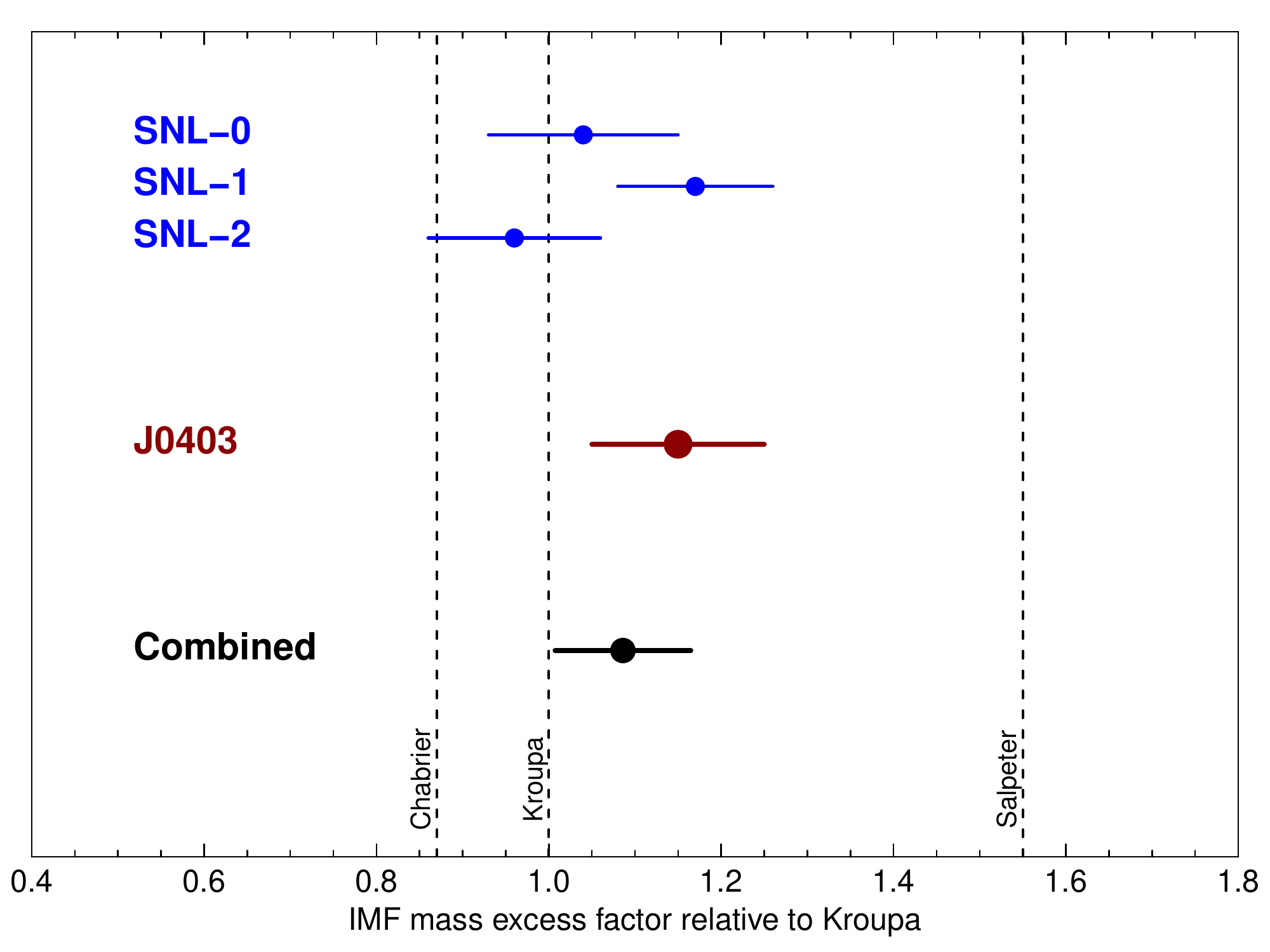}
	\caption{The distribution of the mass excess parameter ($\alpha$) for the combined SNELLS and \galaxy\ sample. In blue, is the SNELLS sample with results for SNL-0 from \citet{Newman2017}, and SNL-1, SNL-2 from \citet{Collier2018}. The result for \galaxy\ from this paper is marked in red. The sample average is $\langle\alpha\rangle$\,=\,1.09\,$\pm$\,0.08, with an inferred {\it intrinsic} scatter of $<$\,0.32, at 90 per cent confidence. These galaxies on average favour a Milky-Way like IMF in preference to a Salpeter or heavier IMF.}
	\label{fig:alphascat}
\end{figure}

We have discovered a new low-redshift gravitational lens, with a bright background galaxy, and a small Einstein radius, probing the stellar-dominated core of a massive elliptical galaxy. We have measured the lensing mass and the $i$--band Einstein aperture luminosity to estimate the IMF mismatch parameter, $\alpha$\,=\,1.15\,$\pm$\,0.10. We attribute a further uncertainty of ($\sim$15 per cent) $\pm$\,0.17 for the systematics. This measurement hence favours a lightweight (MW-like) IMF, rather than a heavy (e.g. Salpeter) one.




In comparison to the SNELLS systems, \galaxy\ is more distant ($z_{\rm lens}$\,=\,0.066 vs 0.031--0.052), but the background source is at much lower redshift ($z_{\rm src}$\,=\,0.19 vs 0.93--2.14). As a result, the Einstein radius is smaller in angular terms (1.5 vs 2.2--2.9 arcsec). Thus despite the greater lens distance, $R_{\rm Ein}$ projects to a similar radius in physical units (1.8 kpc vs 1.5--2.2 kpc) or in galaxy scale units (0.25\,$R_{\rm eff}$ vs 0.3--0.7\,$R_{\rm eff}$). Like the SNELLS galaxies, \galaxy\ has high velocity dispersion and metal abundances typical for massive elliptical galaxies. Furthermore, the lensing aperture mass is compatible with the SNELLS results, favouring a MW-like IMF, and inconsistent with very bottom-heavy IMFs.

Combining our estimate of $\alpha$ for J0403 with the three estimates from SNELLS \citep[taking the values from][i.e. SNL-0, SNL-1, SNL-2\,=\,1.05\,$\pm$\,0.09, 1.17\,$\pm$\,0.09, 0.96\,$\pm$\,0.10]{Newman2017,Collier2018}, as shown in Figure \ref{fig:alphascat}, we can infer limits on the {\it intrinsic} distribution of this quantity among $\sigma$\,$\approx$\,300\,km\,s$^{-1}$ ETGs. We compute the joint likelihood of the four $\alpha$ measurements as a function of the unknown population mean $\langle\alpha\rangle$ and dispersion $\sigma_{\rm int}$, accounting for the measurement errors. Marginalising over $\sigma_{\rm int}$, with a flat prior, we infer $\langle\alpha\rangle$\,=\,1.09\,$\pm$\,0.08 (the error is larger than for a calculation which assumes no intrinsic dispersion). Marginalising over the mean, we infer only an upper limit on the scatter, with $\sigma_{\rm int}$\,$<$\,0.32 at 90 per cent confidence. (For comparison the same treatment applied to SNELLS alone yields an upper limit of $\sigma_{\rm int}$\,$<$\,0.7, highlighting the impact of adding just one new measurement to the analysis.)

One important difference between \galaxy{} and the SNELLS sample, which is relevant for future work, is that the \galaxy\ arcs are both very bright and appear to be quite extended. Hence among all of the known low-redshift lenses, this system is uniquely suitable for pixelized lens inversion methods, which can yield much more powerful constraints on the mass distribution \citep[e.g.][]{Nightingale2017,Oldham2017}. High resolution observations, e.g. with HST, will be essential for such studies.

J0403 is by far the brightest lens system so far identified from our MUSE archival study. The results from the full sample, which is sensitive to much fainter background sources, will be the subject of a future paper (Collier et al. in prep). We have also begun a programme to acquire targeted MUSE observations for selected high-likelihood lenses. The discovery of the \galaxy\ system bodes well for the success of such searches. Together with efforts elsewhere \citep[e.g.][]{Talbot2018}, we anticipate exploiting meaningful samples of lenses at $z$\,$\la$\,0.1 in the near future, to address the continuing puzzle of IMF variations in massive elliptical galaxies.


\section*{Acknowledgements}

W. Collier was supported by an STFC studentship (ST/N50404X/1). RJS and JRL are supported by the STFC Durham Astronomy Consolidated Grant (ST/P000541/1).  This research has made use of the NASA/IPAC Extragalactic Database (NED) which is operated by the Jet Propulsion Laboratory, California Institute of Technology, under contract with the National Aeronautics and Space Administration. This work is based on observations collected at the European Organisation for Astronomical Research in the Southern Hemisphere under ESO programme 098.D-0115(A), retrieved through the ESO Science Archive Facility. The Pan-STARRS1 Surveys (PS1) and the PS1 public science archive have been made possible through contributions by the Institute for Astronomy, the University of Hawaii, the Pan-STARRS Project Office, the Max-Planck Society and its participating institutes, the Max Planck Institute for Astronomy, Heidelberg and the Max Planck Institute for Extraterrestrial Physics, Garching, The Johns Hopkins University, Durham University, the University of Edinburgh, the Queen's University Belfast, the Harvard-Smithsonian Center for Astrophysics, the Las Cumbres Observatory Global Telescope Network Incorporated, the National Central University of Taiwan, the Space Telescope Science Institute, the National Aeronautics and Space Administration under Grant No. NNX08AR22G issued through the Planetary Science Division of the NASA Science Mission Directorate, the National Science Foundation Grant No. AST-1238877, the University of Maryland, Eotvos Lorand University (ELTE), the Los Alamos National Laboratory, and the Gordon and Betty Moore Foundation.





\bibliographystyle{mnras}
\bibliography{newlens.bib} 



%
%
%
%
%
\bsp	
\label{lastpage}
\end{document}